\documentclass[useAMS,usenatbib]{mn2e}
\usepackage{amssymb}

\title[Photometric scaling relations]{Hubble Space Telescope Survey of the Perseus Cluster:~II.~Photometric scaling relations in different environments}
\author[S. De Rijcke, S. Penny, C. Conselice, S. Valcke, E. V. Held]{S. De
Rijcke$^1$, S. J. Penny$^2$, C. J. Conselice$^2$, S. Valcke$^1$,
E. V. Held$^3$ \\ $^1$ Sterrenkundig Observatorium, Universiteit Gent,
Krijgslaan 281, S9, B-9000, Gent, Belgium\\ $^2$ School of Physics \&
Astronomy, University of Nottingham, Nottingham NG7 2RD, UK \\ $^3$
Osservatorio Astronomico di Padova, INAF, vicolo dell'Osservatorio 5,
35122 Padova, Italy}

\begin{document}

\date{Accepted 1988 December 15. Received 1988 December 14; in
original form 1988 October 11}

\pagerange{\pageref{firstpage}--\pageref{lastpage}} \pubyear{2007}

\maketitle

\label{firstpage}

\begin{abstract}
We investigate the global photometric scaling relations traced by
early-type galaxies in different environments, ranging from dwarf
spheroidals, over dwarf elliptical galaxies, up to giant ellipticals
($-8\,{\rm mag} \gtrsim {\rm M}_V \gtrsim -24$~mag). These results are
based in part on our new HST/ACS F555W and F814W imagery of dwarf
spheroidal galaxies in the Perseus Cluster. The full sample, built
from our HST images and from data taken from the literature, comprises
galaxies residing in the Local Group; the Perseus, Antlia, Virgo, and
Fornax Clusters; and the NGC5898 and NGC5504 groups.

Photometric parameters, such as the half-light radius, the central
surface brightness, and the S\'ersic exponent $n$ are used to
parameterize the light distributions and sizes of early-type
galaxies. These parameters all vary in a continuous fashion with
galaxy luminosity over a range of more than six orders of magnitude in
luminosity. We also find that all early-type galaxies follow a single
color-magnitude relation, which we interpret as a
luminosity-metallicity relation for old stellar populations. These
scaling relations are almost independent of environment, with Local
Group and cluster galaxies coinciding in the various diagrams. As an
example, due the presence of a population of very low surface
brightness dSphs in the Fornax cluster, which may be tidally heated
dwarf galaxies, the Fornax dSph population is on average only
0.2~mag~arcsec$^{-2}$ fainter than the Local Group dSph
populations. This offset is much too small to destroy the global
relation between luminosity and central surface brightness.

We show that at ${\rm M}_V \sim -14\,{\rm mag}$, the slopes of the
photometric scaling relations involving the S\'ersic parameters change
significantly. This contradicts previous claims that the relations
involving S\'ersic parameters are pure power-laws for all early-type
galaxies and are, therefore, more fundamental than other photometric
scaling relations derived from them. We argue that these changes in
slope reflect the different physical processes that dominate the
evolution of early-type galaxies in different mass regimes. As such,
these scaling relations contain a wealth of information that can be
used to test models for the formation of early-type galaxies.
\end{abstract}


\section{Introduction}

Dwarf spheroidal galaxies (dSphs) are faint stellar systems (M$_V
\gtrsim -14$~mag) with smooth elliptical isophotes. They are presumed
to be the faint analogs of dwarf elliptical galaxies (dEs), which are
usually defined as lying in the luminosity range $-19\,{\rm mag}
\lesssim {\rm M}_V \lesssim -14$~mag. Being found typically not more
than a few hundred kiloparsecs away from a massive galaxy or in groups
and clusters of galaxies, they show a strong predilection for
high-density environments \citep{m98,ggh03}. Their dynamical
mass-to-light ratios, derived by fitting dynamical models to their
stellar velocity dispersion profiles or based on stability arguments,
vary from a few tens up to a few hundreds, in solar units
\citep{m98,l02,k05,de06,l07,m08,pe08}. This high mass-to-light ratio
suggests the presence of copious amounts of dark matter that help
protect the embedded stellar body of the dSph against the tidal forces
of the massive host galaxy or of the galaxy cluster or group in which
the dSph resides. This would explain why only a handful of dSphs in
the Local Group show clear signs of an ongoing interaction despite
being close satellites of either the Milky Way or M31
\citep{jsh95,m96,mi06,s07,l07}. Many dSphs and dEs still contain an
interstellar medium and some even host low-level star formation
\citep{br00,ggh03,cogw03,de03,b05,l06,y07}, showing that supernova
explosions are not very efficient at expelling gas
\citep{mbe03,vdd08}.

Bright elliptical galaxies, or Es, and dEs follow the same photometric
and kinematic scaling relations \citep{gg03,mg05,de05,smc08}. In the
luminosity interval $-24\,\,{\rm mag} \lesssim {\rm M}_V \lesssim
-14$~mag the parameters of the S\'ersic profile follow simple
power-laws as a function of luminosity and early and late type
galaxies trace parallel Tully-Fisher relations \citep{de07}. From this
wealth of data a picture of (dwarf) galaxy formation emerges that
suggests an underlying unity in the physics driving the formation and
evolution of stellar systems, with the environment playing a role that
is in many situations subordinate to that of internal processes. More
specifically, numerical simulations and semi-analytic models of galaxy
formation within a $\Lambda$CDM cosmology can account for the observed
scaling relations when taking into account supernova feedback in
galactic gravitational potential wells steepening with galaxy mass
\citep{c01,ny04,ri05,m06,vdd08}.

The goal of this paper is to investigate whether the photometric
scaling relations traced by dEs and Es persist down to the dSphs
(${\rm M}_V \gtrsim -14$~mag). We also check for possible
environmental influences, other than the obvious density-morphology
relation (i.e. the fact that dEs/dSphs are found predominantly in
high-density environments). We present new data based on our HST/ACS
imaging of dSphs/dEs in the Perseus Cluster and combine these with
data of early-type galaxies in the Antlia, Fornax, and Virgo Clusters
and the Local Group (see section \ref{data}). The resulting
photometric scaling laws are presented in section \ref{dare}. We
discuss the results in section \ref{disc}.

\section{Photometric data} \label{data}

\begin{figure*}
\vspace*{14.5cm}
\special{hscale=80 vscale=80 hsize=720 vsize=450
hoffset=0 voffset=-190 angle=0 psfile="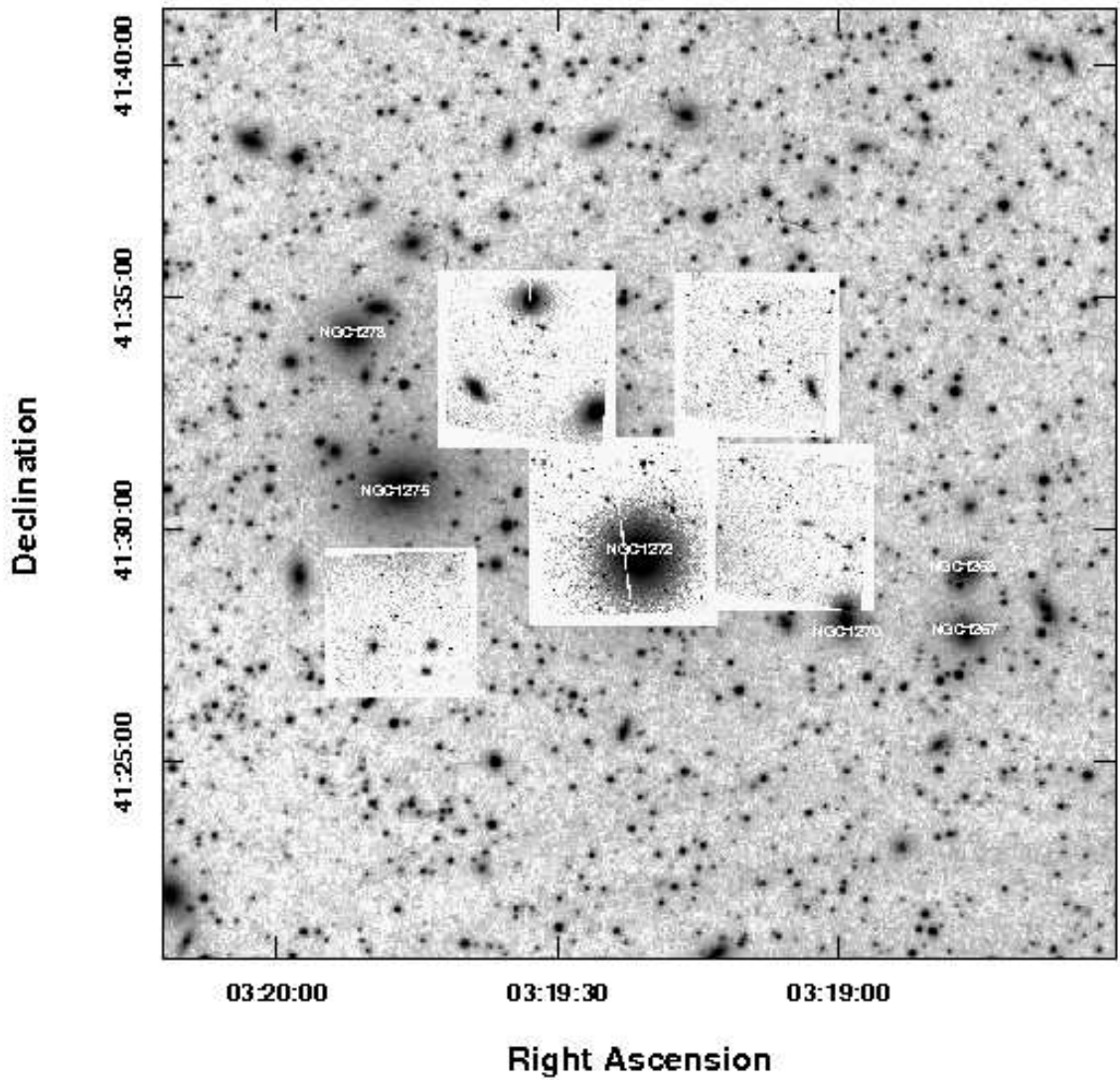"}
\caption{Position on the sky of our five HST/ACS pointings near the
core of the Perseus cluster. The HST/ACS fields are overplotted onto a
DSS image of the cluster. The NGC-numbers of the most prominent
cluster members are indicated in the figure. \label{pointings}}
\end{figure*}

\subsection{Perseus Cluster data}

The Perseus Cluster (Abell 426) is one of the richest nearby galaxy
clusters, with a redshift $v$ = 5366~km s$^{-1}$ \citep{Stublerood99},
and at a distance $D$ = 72~Mpc (as given by NED). Due to its low
Galactic latitude ($b$ $\approx$ $-13^{\circ}$) it has not been
studied in as much detail as other nearby clusters such as Fornax,
Virgo and Coma. We have obtained high resolution \textit{Hubble Space
  Telescope (HST)} Advanced Camera for Surveys (ACS) WFC imaging in
the F555W and F814W bands of five fields in the Perseus Cluster core,
in the immediate vicinity of NGC1275 and NGC1272, the cluster's
brightest members, obtained in 2005 (program GO 10201). The scale of
the images is 0.05$''$ pixel$^{-1}$, with a field of view of $202''
\times 202''$, providing a total survey area of $\sim$ 57
arcmin$^{2}$. The positions on the sky of these five pointings are
presented in Fig. \ref{pointings}. Exposure times were 2368 and 2260
seconds for the F555W and F814W bands, respectively. The fields were
chosen to cover the most likely cluster dSphs and dEs identified from
ground-based imagery by \cite{cgw03}. For some of these, there is
spectroscopic confirmation of their cluster membership
\citep{pc08}. For the others, we use morphological criteria to decide
cluster membership. The CAS system for quantifying compactness,
asymmetry, and clumpiness/smoothness \citep{c03} proves very useful to
reject e.g. background spiral galaxies based on a smoothness criterion
and background bright ellipticals based on a compactness criterion
(see \cite{pe08}). The Perseus dataset straddles the dE-dSph
transition at M$_V \sim -14$~mag and is therefore essential to the
discussion that follows.

\begin{figure*}
\vspace*{8cm}
\special{hscale=50 vscale=50 hsize=250 vsize=250
hoffset=-28 voffset=-130 angle=0 psfile="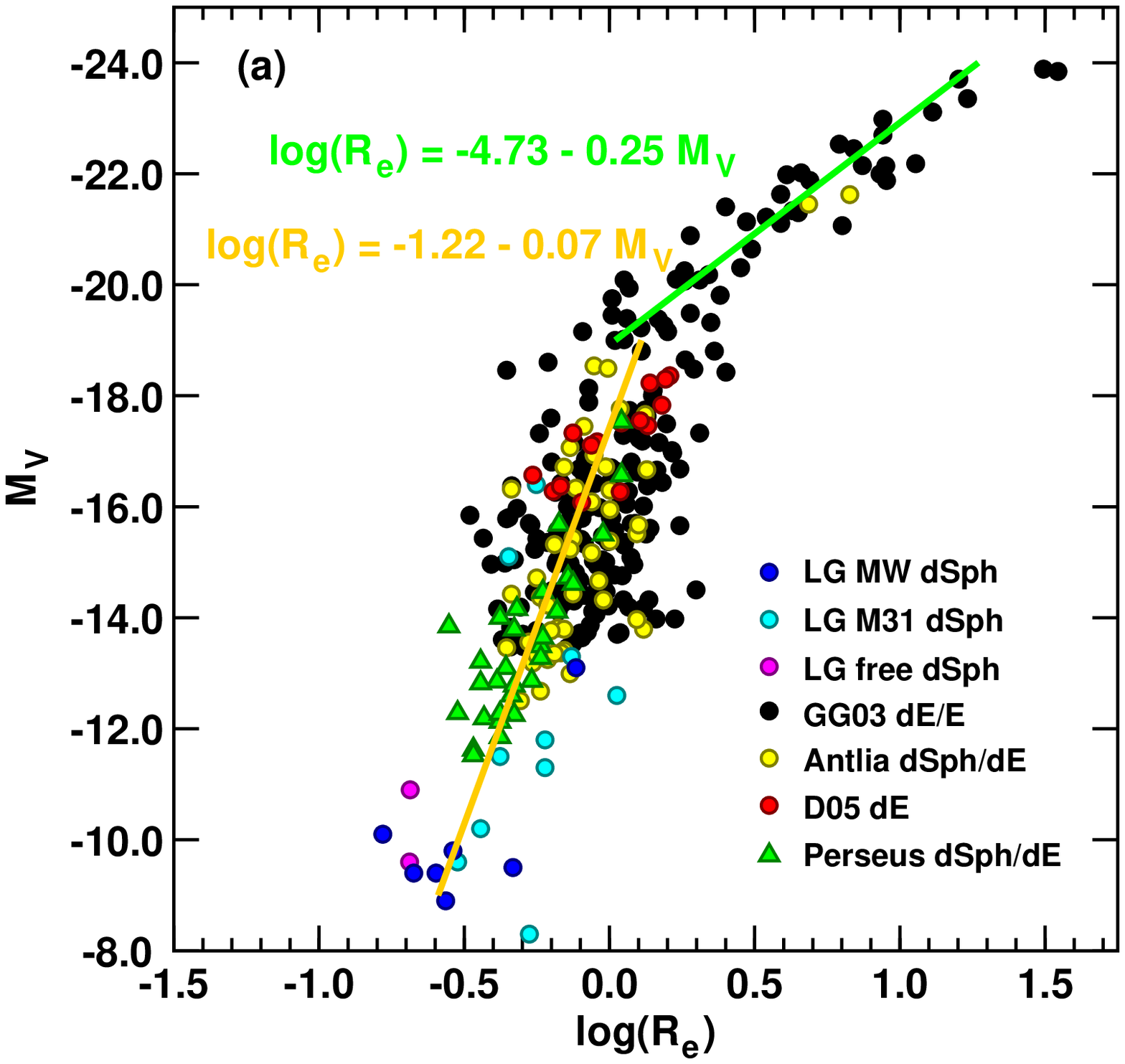"}
\special{hscale=50 vscale=50 hsize=550 vsize=250
hoffset=230 voffset=-130 angle=0 psfile="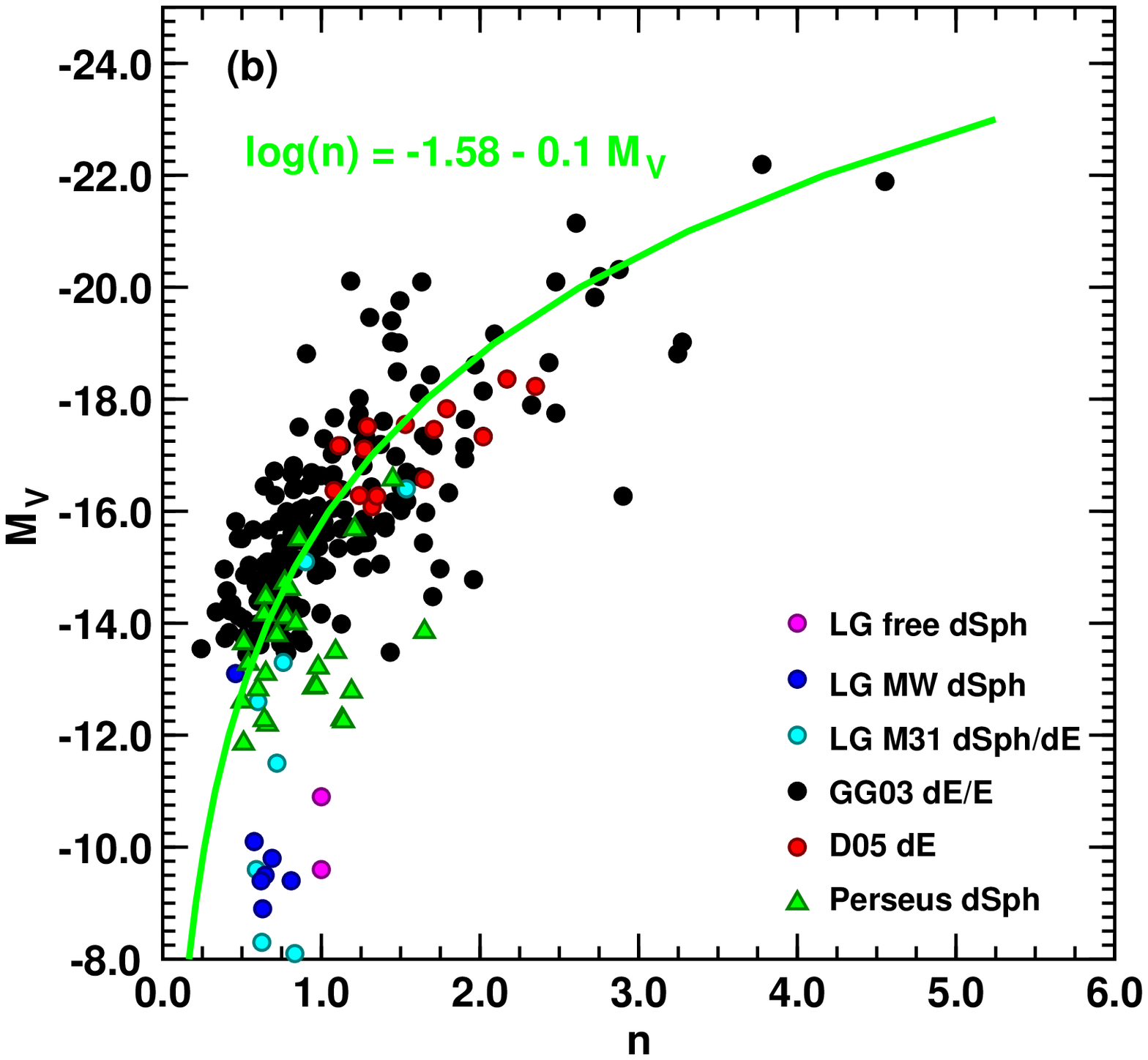"}
\vspace*{8.1cm}
\special{hscale=50 vscale=50 hsize=250 vsize=250
hoffset=-28 voffset=-130 angle=0 psfile="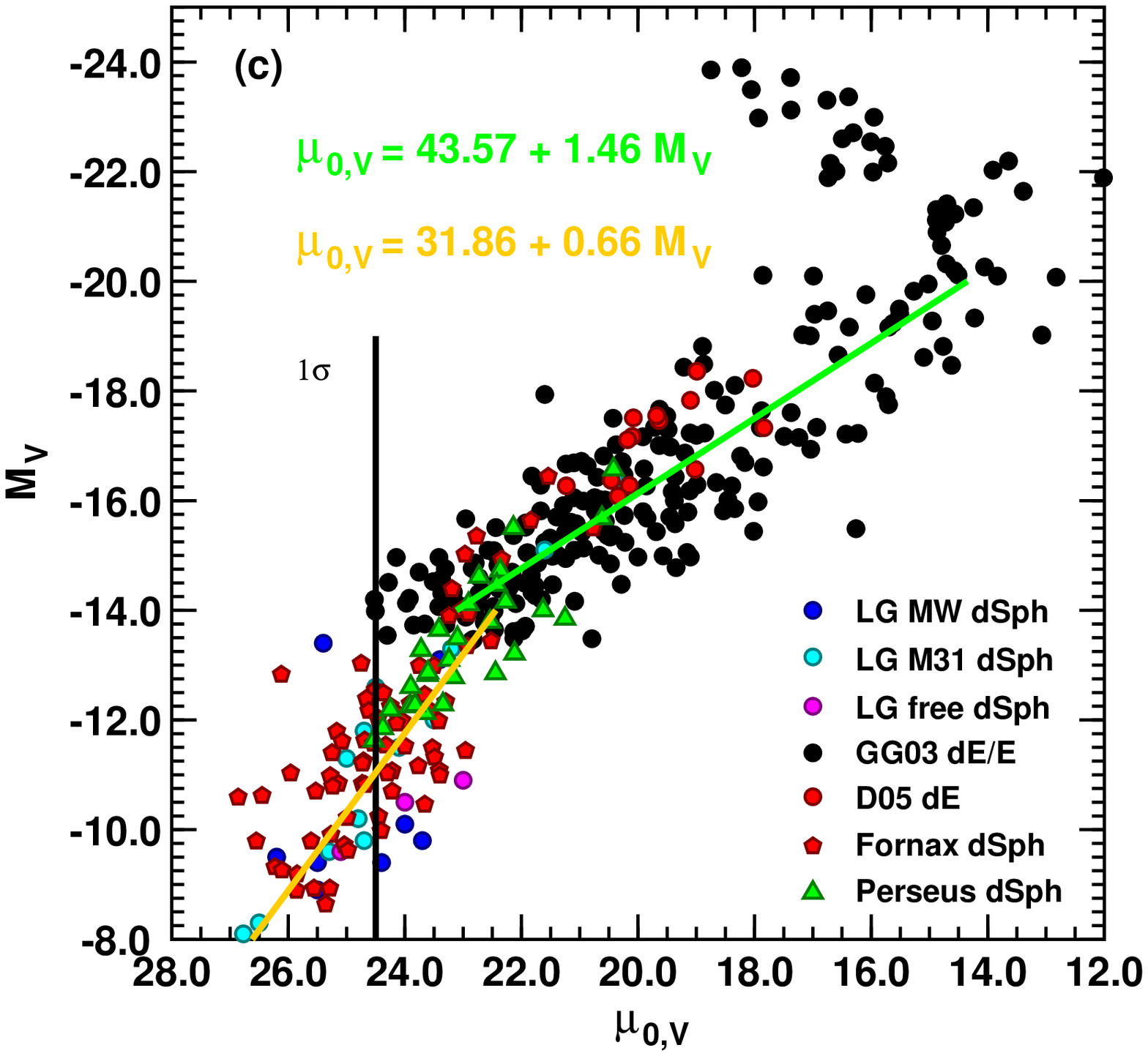"}
\special{hscale=50 vscale=50 hsize=550 vsize=270
hoffset=230 voffset=-130 angle=0 psfile="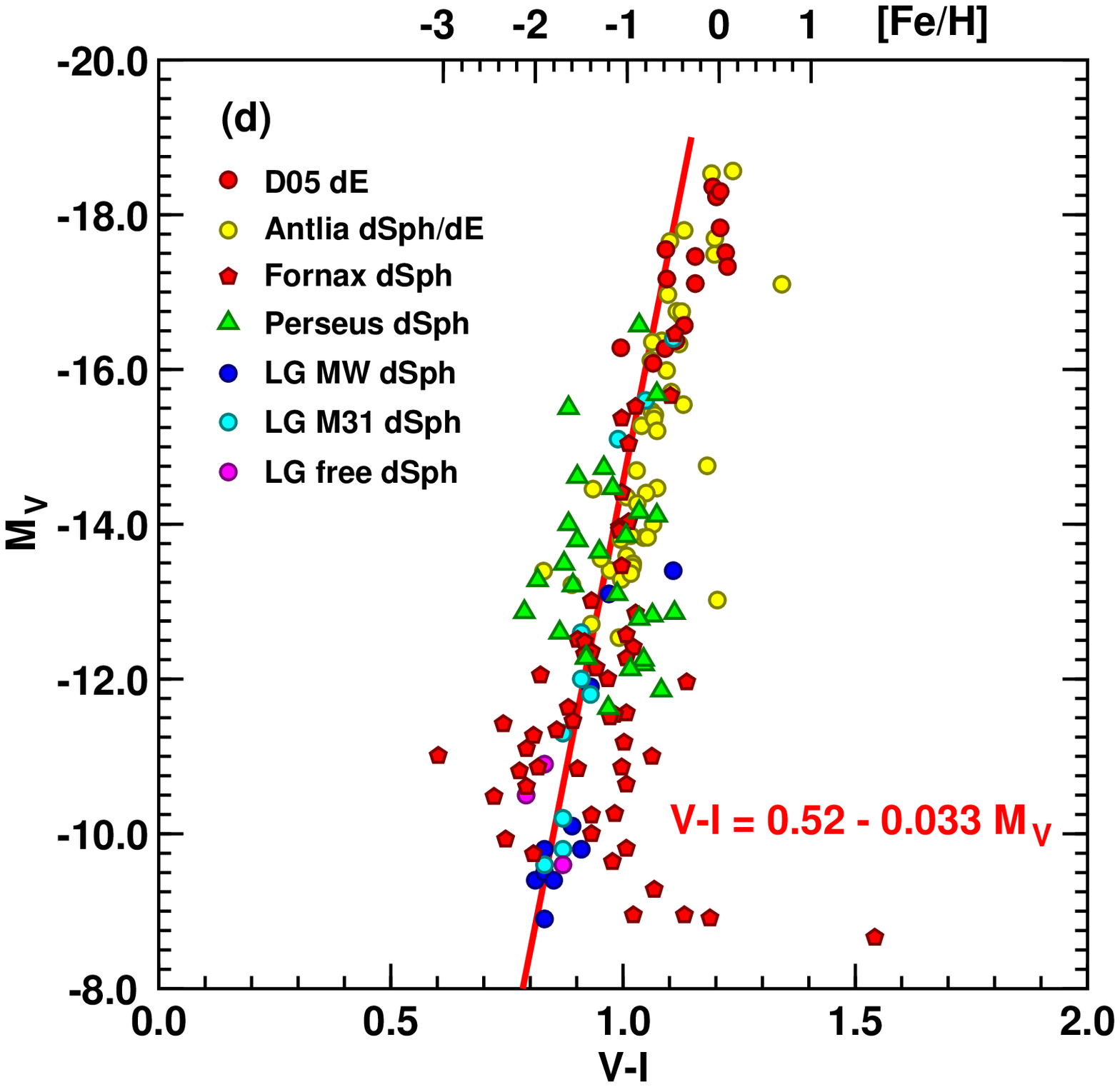"}
\caption{Photometric scaling relations of dwarf spheroidal, dwarf
  elliptical, and elliptical galaxies. {\bf (a)} Half-light radius,
  $R_{\rm e}$ (kpc), versus absolute V-band magnitude, ${\rm
    M}_V$. The green line traces the relation $\log(R_{\rm e}) = -4.73
  - 0.25 \times {\rm M}_V$ of the bright ellipticals whereas the
  orange line indicates the relation $\log(R_{\rm e}) = -1.22 - 0.07
  \times {\rm M}_V$ of the dEs and dSpshs. {\bf (b)} The S\'ersic
  parameter $n$ versus absolute V-band magnitude. The green curve
  traces the $\log(n) \propto 0.1 \times {\rm M}_V$ relation of Graham
  \& Guzm\'an (2003). {\bf (c)} Central surface brightness $\mu_{0,V}$
  (mag~arcsec$^{-2}$), derived by fitting a S\'ersic law to the
  observed galaxies' surface brightness profiles, versus absolute
  V-band magnitude. The green curve traces the relation $\mu_{0,V}
  \propto 1.46 \times {\rm M}_V$ of the dwarf and giant ellipticals;
  the orange curve is the $\mu_{0,V} \propto 0.66 \times {\rm M}_V$
  relation of the dwarf spheroidals. {\bf (d)} V$-$I colour versus
  absolute V-band magnitude (the top $x$-axis is labeled in
  metallicity, quantified by [Fe/H], using an empirical relation
  between V$-$I and [Fe/H] (Couture, Harris, Allwright (1990); Kundu
  \& Whitmore (1998)). The red line traces the linear fit to the
  Fornax cluster M$_V$ vs. V$-$I relation by Mieske et al. (2007). The
  photometric data of the Local Group dSphs ({\sf LG MW dSph} for the
  Milky Way satellites; {\sf LG M31 dSph/dE} for the M31 satellites;
  {\sf LG free dSph} for the dSphs that are not linked to a giant host
  galaxy) come from \protect\cite{p93}, \protect\cite{ih95},
  \protect\cite{sv96}, \protect\cite{c99}, \protect\cite{ggh03},
  \protect\cite{mi06}, \protect\cite{mai07}, \protect\cite{z07}. The
  other data sets are from \protect\cite{gg03} ({\sf GG03 dE/E}),
  \protect\cite{de05} ({\sf D05 dE}), \protect\cite{mi07} ({\sf Fornax
    dSph}), \protect\cite{smc08} ({\sf Antlia dSph/dE}) and this work
  ({\sf Perseus dSph}).
\label{LVVI}}
\end{figure*}

A detailed account of our photometric analysis of the HST/ACS images
of Perseus dSphs and of the properties of the individual galaxies will
be reported elsewhere (Penny, Conselice, De Rijcke, Held, in prep.). 

\subsection{Data from the literature}

The photometric data, including resolved photometry for surface
brightness profiles, of the Local Group dSphs that are identified as
Milky Way satellites are collected from \cite{ggh03} and \cite{ih95},
adopting the distances listed in \cite{ggh03}. Data of the M31 dSph
satellites is taken from \cite{p93}, \cite{c99}, \cite{ggh03},
\cite{mi06}, \cite{mai07}, and \cite{z07}. Data of three Local Group
dSphs that are not linked to a giant host galaxy, the Tucana dSph,
DDO210, and KKR25, come from \cite{sv96}, \cite{ggh03}, and
\cite{mi06}. \cite{de05} (D05) and \cite{mi07} provide photometric
data on the early-type dwarf galaxy population of the Fornax
cluster. Half of the D05 sample consists of dEs from the NGC5044 and
NGC5989 groups. The data of the dSphs and dEs in the Antlia cluster
are taken from \cite{smc08}. Data data for the giant elliptical and
for Coma dEs is taken from \cite{gg03} (GG03).

This sample of early-type galaxies comprises dwarf spheroidals, with
$-14\,\,{\rm mag} \lesssim {\rm M}_V \lesssim -8$~mag, dwarf
ellipticals, with $-19\,\,{\rm mag} \lesssim {\rm M}_V \lesssim
-14$~mag, and bright ellipticals, with ${\rm M}_V \lesssim
-19$~mag. We plot the positions of the sample galaxies in diagrams of
V-band absolute magnitude vs. {\em (i)} half-light radius R$_{\rm e}$
(in kpc), vs. {\em (ii)} the S\'ersic exponent $n$ of the best fitting
S\'ersic profile, vs. {\em (iii)} the central V-band surface
brightness of the best fitting S\'ersic profile, and vs.  {\em (iv)}
V$-$I colour. These diagrams are shown in Fig. \ref{LVVI}. The
datasets are plotted using different symbols that are explained within
each panel of the figure. The presence of a dataset in a given panel
depends solely on the availability of the required data.

\section{Photometric scaling relations} \label{dare}

\subsection{Method}

For our Perseus dSphs/dEs, we measure the profiles of
surface-brightness, position angle, and ellipticity as a function of
the geometric mean of major and minor axis distance, denoted by $a$
and $b$ respectively, using our own software. Basically, the code fits
an ellipse through a set of positions where a given surface brightness
level is reached. Residual cosmics, background galaxies, and
foreground stars are masked and not used in the fit. The shape of an
isophote, relative to the best fitting ellipse, is quantified by
expanding the surface brightness variation along this ellipse in a
fourth order Fourier series with coefficients $S_4$, $S_3$, $C_4$, and
$C_3$:
\begin{eqnarray}
I(a,\theta) &=& I(a) \left[ 1 + C_3(a) \cos(3\theta)+ C_4(a)
\cos(4\theta) \right. \nonumber \\
&& \left.+ S_3(a)\sin(3\theta)+
S_4(a) \sin(4\theta) \right]
\end{eqnarray}
Here, $I(a)$ is the mean surface brightness of an isophote with
semi-major axis $a$, and the angle $\theta$ is measured from the major
axis. Apparent ABMAG magnitudes in the F555W and F814W bands are
calculated using the zero-points given by \citet{si05}. These
magnitudes are corrected for interstellar reddening adopting the color
excess E(B$-$V)$\,=0.171$~mag \citep{sch98} and using the
prescriptions given in \citet{si05}. These reddening-corrected
magnitudes are finally converted into Johnson V and I band magnitudes
using the transformations of \citet{si05}.

The smooth representation of a galaxy's surface-brightness profile,
$I(a,\theta)$, is then subtracted from the original image. We have
checked that the result is indeed a pure noise image, free of
residuals. $I(a,\theta)$ is integrated over circular apertures out to
the last isophote we could reliably measure (which is at $\mu_{\rm
  ABMAG} \approx 27$~mag~arcsec$^{-2}$ in both the F555W and F814W
images) to derive model independent structural parameters, such as the
total apparent magnitude and the half-light radius in each band. For
such deep images of galaxies with a roughly exponentially declining
surface brightness profile, this truncation results in an
insignificant uncertainty on the total luminosity, of the order of a
few per cent (see also \cite{de05}). V$-$I colors are measured using
the V and I-band flux inside the I-band half-light radius. We fit a
S\'ersic profile, given by
\begin{equation}
\mu_V(r_p) = \mu_{0,V} + 1.0857\left( \frac{r_p}{r_0} \right)^{1/n},
\label{sereq}
\end{equation}
to the V-band surface brightness profiles of the program galaxies,
expressed in mag~arcsec$^{-2}$. Here, the equivalent radius $r_p =
\sqrt{ab}$ is the geometric mean of the isophotes' semi-major axes $a$
and semi-minor axes $b$, $\mu_{0,V}$ is the central V-band surface
brightness, in mag~arcsec$^{-2}$, $r_0$ is a scale-length, in
arcseconds, and the exponent $n$ is a shape parameter with $n=1$
giving an exponentially declining profile and $n=4$ corresponding to
the de Vaucouleurs-profile typical of giant ellipticals.

For the Antlia cluster, we adopt a distance of 35.1~Mpc, as advocated
by \cite{smc08}. This distance estimate is based on surface-brightness
fluctuations (SBF) distances to two giant Es in this cluster. As in
D05, we place the Fornax cluster at a distance of 19.7 Mpc, the NGC
5044 group at 35.1 Mpc, the NGC 5898 group at 30.3 Mpc, and the NGC
3258 group at 40.7 Mpc, all in good agreement with SBF distances
\citep{t01,j03}. For the GG03 data set, only B-band photometry is
available. We convert B-band magnitudes into V-band magnitudes using a
B$-$V color-magnitude relation constructed from the ${\rm M}_V-({\rm
  V}-{\rm I})-$[Fe/H] relation (see section \ref{subLVVI}) in
combination with SSP models for 10~Gyr old stellar populations
\citep{vz96}. This relation interpolates between B$-$V$\,\approx
0.7$~mag at M$_B=-8$~mag and B$-$V$\,\,\approx 1.0$~mag at
M$_B=-22$~mag. As we show below, our conclusions do not depend on this
slight color correction applied to the GG03 dataset. There are no
systematic deviations of the color corrected GG03 dataset with respect
to other datasets with which it overlaps in luminosity. Applying a
constant mean color correction $\langle {\rm B}-{\rm V} \rangle =
0.8$~mag \citep{vzb04} yields essentially the same results.

For the Local Group dSphs for which no S\'ersic parameters can be
found in the literature, we fit S\'ersic profiles, with an added
constant background density of stars, to the star counts of the dSphs
presented in \cite{ih95}.

We now place these early-type galaxies in diagrams correlating the
V-band absolute magnitude M$_V$, the S\'ersic exponent $n$, the
extrapolated central surface brightness $\mu_{0,V}$, and the V$-$I
colour. The goal is to investigate the behaviour of the relations
between these structural parameters as a function of luminosity in the
range $-24\,{\rm mag} < {\rm M}_V < -8$~mag and of environment, using
galaxies from the Local Group, the NGC5044 and NGC5898 groups, and the
Antlia, Fornax, Perseus, and Coma clusters.

\subsection{Luminosity vs. half-light radius}

\begin{figure}
\vspace*{6cm}
\special{hscale=40 vscale=40 hsize=720 vsize=180
hoffset=-20 voffset=220 angle=-90 psfile="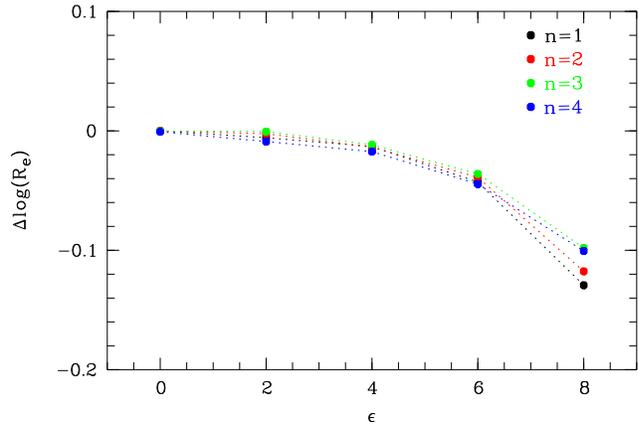"}
\caption{Deviation of the logarithm of the half-light radius measured
using circular apertures from the half-light radius defined as the
geometric mean of the semi-major axis $a$ and semi-minor axis $b$ of
the elliptical isophote that encloses half the light as a function of
flattening $\epsilon = 10(1-b/a)$. This exercise was performed for
synthetic surface brightness profiles with S\'ersic $n=1$, 2, 3, and
4. The maximum deviation is not larger than $\Delta \log({\rm R}_{\rm e})
\approx 0.15$. \label{fitre}}
\end{figure}

We plot V-band absolute magnitude against half-light radius, denoted
by $R_{\rm e}$, in panel {\bf (a)} of Fig. \ref{LVVI}. Es and gEs
follow a trend of increasing half-light radius with increasing
luminosity, which can be quantified as
\begin{equation}
\log(R_{\rm e}) = (-4.73 \pm 0.47) - (0.25 \pm 0.02) \times {\rm M}_V
\end{equation}
between M$_V \sim -19$~mag and M$_V \sim -24$~mag (green line in panel
{\bf (a)} of Fig. \ref{LVVI}). In the range $-19\,{\rm mag} \lesssim
{\rm M}_V \lesssim -14$~mag, on the other hand, the half-light radius
of dEs increases much more slowly as a function of luminosity, with
\begin{equation}
\log(R_{\rm e}) = (0.92 \pm 0.14) - (0.05 \pm 0.01) \times {\rm M}_V
\end{equation}
(see also \cite{gg03,de05,smc08,gw08}). In the very low-luminosity
dSph regime, radius again appears to increase slightly more rapidly
with luminosity as
\begin{equation}
\log(R_{\rm e}) = (-1.40 \pm 0.16) - (0.09 \pm 0.01) \times {\rm M}_V,
\end{equation}
between M$_V \sim -8$~mag and M$_V \sim -14$~mag. This trend appears
to continue for even fainter Milky Way satellites \citep{mdr08}. The
very gentle slope change around M$_V \sim -14$~mag seems to be caused
mostly by the GG03 data set. All other datasets rather suggest that
half-light radius behaves as a power-law as a function of luminosity
in the regime M$_V \gtrsim -19$~mag. A fit to the data of the dwarf
galaxies fainter than M$_V = -19$~mag from the D05, Antlia, Perseus
and Local Group datasets gives
\begin{equation}
\log(R_{\rm e}) = (-1.22 \pm 0.11) - (0.07 \pm 0.01) \times {\rm M}_V
\end{equation}
(orange line in panel {\bf (a)} of Fig. \ref{LVVI}). The gentle
curvature of the M$_V-$R$_{\rm e}$ relation around M$_V \sim -19$~mag
was shown by \cite{gg03} to be a consequence of the power-law
dependence of the S\'ersic parameters on galaxy luminosity.

One small caveat, however:~the half-light radius can be measured in a
number of different ways which do not necessarily yield the same
result for a given galaxy. One can, for instance, use the geometric
mean of the semi-major axis and semi-minor axis of the elliptical
isophote that encloses half the light as a measure for R$_{\rm e}$. Or one
can fit a S\'ersic law to the surface brightness profile, evaluated as
a function of equivalent radius, and adopt the half-light radius of
this model profile, as in \cite{gg03}. Or one can construct the growth
curve model-independently by integrating the observed surface
brightness over circular apertures and thus derive a half-light
radius, as in \cite{de05}. For spherically symmetric galaxies, these
different approaches obviously yield the same result. For very
flattened galaxies, they may differ. If the early-type galaxy
population shows a trend of mean flattening with luminosity (see e.g.
\citet{b89}), this might introduce a spurious trend of half-light
radius with luminosity.

\begin{figure*}
\vspace*{14.5cm}
\special{hscale=65 vscale=65 hsize=550 vsize=420
hoffset=0 voffset=550 angle=-90 psfile="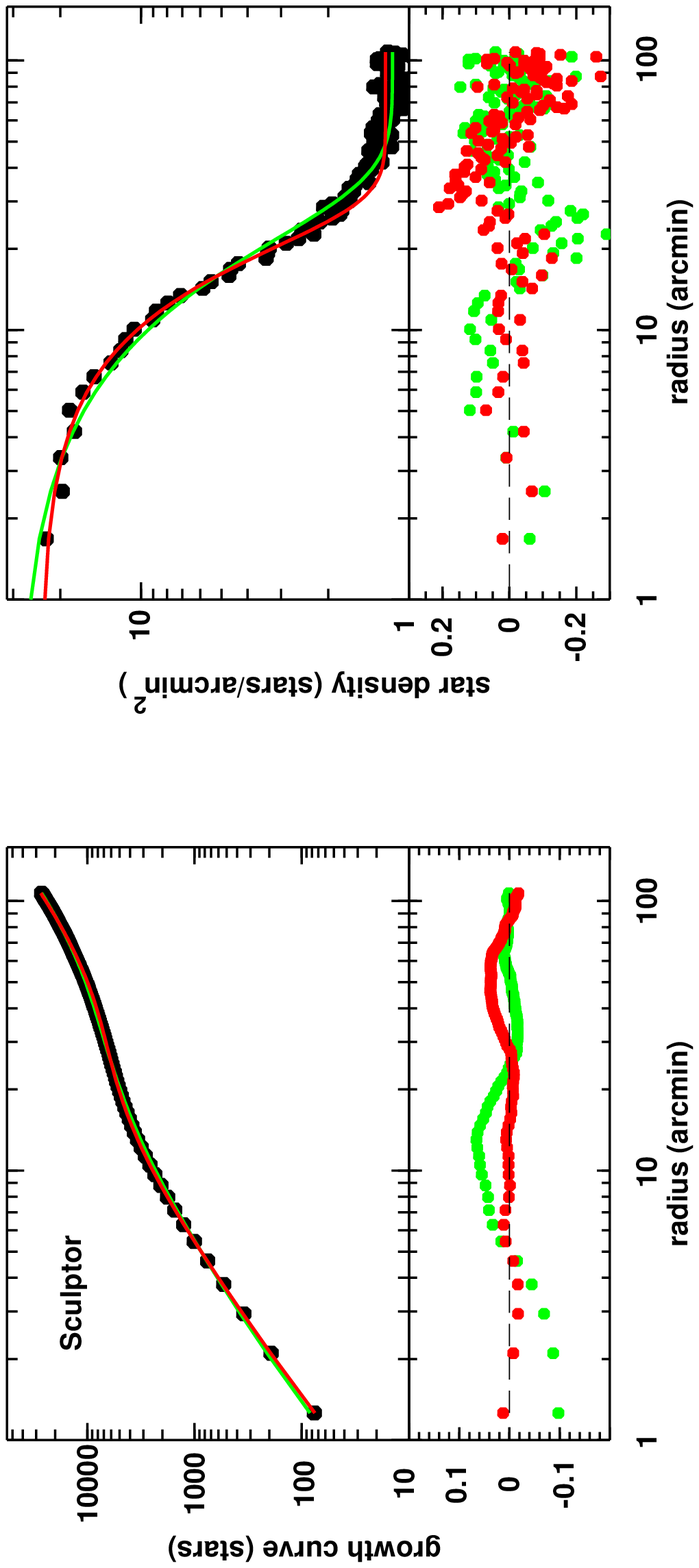"}
\special{hscale=65 vscale=65 hsize=550 vsize=210
hoffset=0 voffset=340 angle=-90 psfile="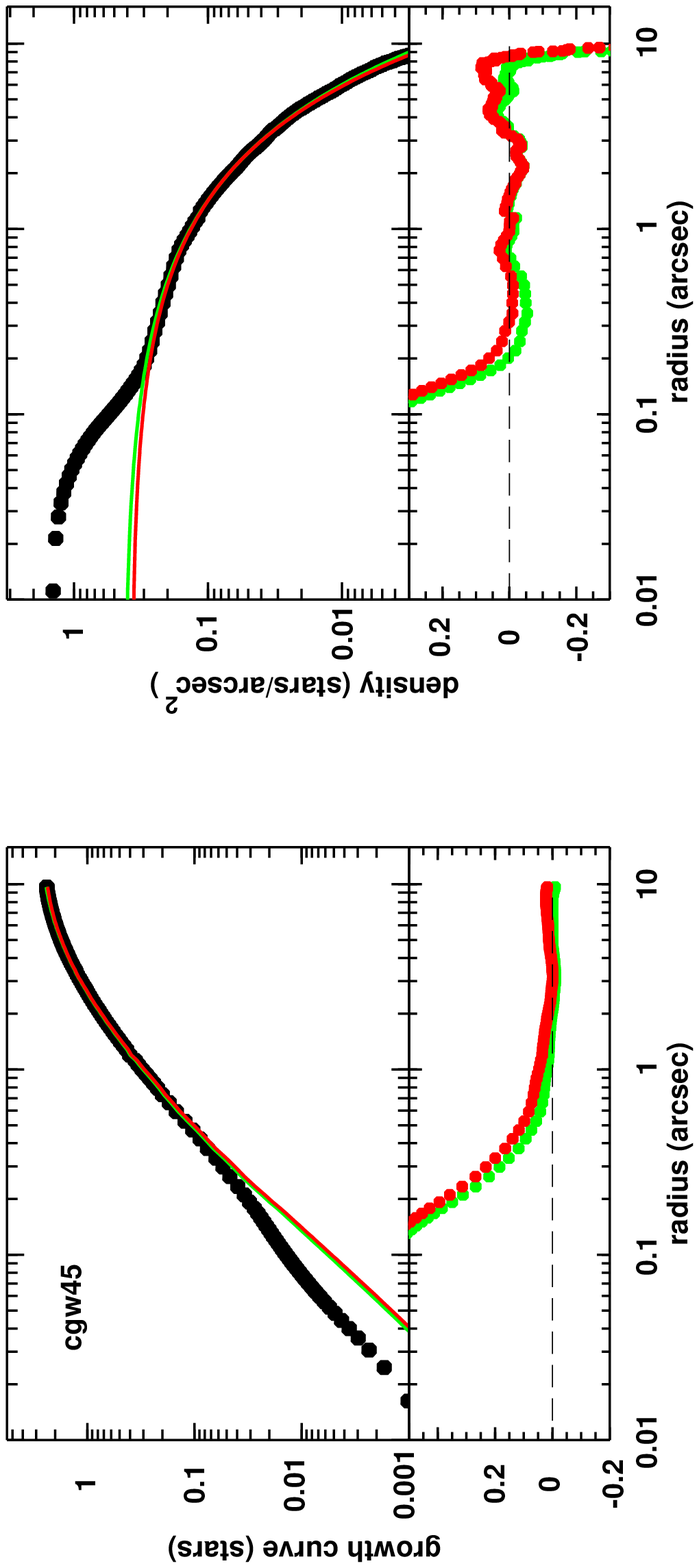"}
\caption{The growth curve (left panels) and the radial profile of the
  projected stellar density (right panels) of the Sculptor dSph (top
  row), taken from Irwin \& Hatzidimitriou (1995), and the Perseus
  cluster dSph CGW45 (bottom row), taken from the Conselice,
  Gallagher, Wyse (2003) sample . The stellar density of CGW45 is in
  arbitrary units. The black dots represent the data; the red line is
  a S\'ersic profile fitted to the projected stellar density profile;
  the green line is a S\'ersic profile fitted to the growth
  curve. Under each panel, the relative residuals are plotted in the
  same colours. Sculptor is an example of a galaxy whose surface
  brightness profile deviates from a S\'ersic profile at large radii;
  CGW45 is a nucleated dwarf elliptical galaxy and deviates from a
  S\'ersic profile at small radii.
\label{ser1}}
\end{figure*}

In order to check how large this systematic effect can be, we create
synthetic galaxy images using flattened S\'ersic profiles with
different flattenings $\epsilon = 10(1-b/a)$ and S\'ersic exponents
$n$. The half-light radius is then measured as the geometric mean of
the semi-major axis and semi-minor axis of the elliptical isophote
that encloses half the light and by integrating over circular
isophotes. The difference between the two approaches is plotted in
Fig. \ref{fitre} as a function of flattening $\epsilon$ and S\'ersic
$n$. The maximum deviation is obtained for small $n$ and strong
flattening and is not larger than $\Delta \log({\rm R}_{\rm e}) \approx
0.15$. This is smaller than the scatter on the observed scaling
relations. We therefore do not expect to see any significant
systematic trend of half-light radius with flattening as a result of
the particular way it was measured.

Although the data show considerable scatter, especially in the dwarf
regime, it is striking that galaxies from a wide range of environments
trace a continuous M$_V$ vs. R$_{\rm e}$ relation over a range of 6 orders
of magnitude in luminosity. 
%Moreover, the N-body/SPH simulations of
%\cite{vdd08} quite faithfully reproduce the luminosity-size relation
%of the dSphs and the break around M$_V \sim -14$~mag.

%Within the context of these calculations, this is
%a consequence of the balance between on the one hand supernova
%explosions, that heat and expell gas by feedback, and the on the other
%hand the gravitational potential, which drives cooling gas towards the
%galaxy center and which steepens with galaxy mass. Roughly three
%distinct episodes of star formation can be identified during the life
%of a model galaxy (see Fig. 7 of \cite{vdd08}). In the least massive
%models, the supernovae win over the shallow gravitational
%potential. Gas is supported against collapse and star formation occurs
%throughout the galaxy. After the first star-formation event, the
%half-light radius remains constant or increases only slightly with
%time. In the most massive models, the steep gravitational potential
%wins over the supernovae. After the first star-formation episode, gas
%cools and sinks towards the galaxy center where subsequent centrally
%concentrated star-formation events take place. In these models, the
%half-light radius decreases with time. 

\subsection{Luminosity vs. S\'ersic $n$}

It is well known that early-type galaxies brighter than M$_V \sim
-14$~mag trace a single M$_V$ vs. $n$ relation. This relation is
quantified as $\log(n) = -1.4 - 0.1 \times {\rm M}_B$ by \citet{jb97},
as $\log(n) = -1.88 - 0.12 \times {\rm M}_{\rm F606W}$ by
\citet{gg03}, and as $\log(n) = -1.52 - 0.11 \times {\rm M}_B$ by
\citet{gw08}. These authors fit S\'ersic profiles to the observed
surface brightness profiles, evaluated as function of equivalent
radius. However, it is also possible to fit the growth curve,
constructed by integrating a galaxy image over circular apertures,
with the growth curve of the S\'ersic profile. This approach is
advocated by \citet{ps97}. The latter method of determining the
S\'ersic exponent $n$ assigns a large weight to the outer data points
and has the obvious advantage of yielding a S\'ersic model with the
same total luminosity as the observed galaxy. However, the model's
surface brightness profile does not necessarily provide an acceptable
fit to that of the observed galaxy. The former method, on the other
hand, assigns a large weight to the inner data points. This method has
the benefits that clearly non-S\'ersic components (such as a nuclear
star cluster of a stellar halo) can be omitted from the fit and that
the model provides a good fit to the observed surface brightness
profile (at least within some specified radial range). It does
not, however, necessarily provide a good approximation of the galaxy's
growth curve.

As an example, we apply both methods to the Sculptor dSph, with star
count data taken \cite{ih95}, and to our new HST/ACS photometry of
CGW45, a Perseus cluster dE, taken from the sample of
\cite{cgw03}. The observed surface brightness profile of CGW45,
$\mu(r)$, is converted into a ``star counts'' profile in arbitrary
units, $n(r)$ using
\begin{equation}
n(r) = 10^{(20-\mu(r))/2.5}.
\end{equation}
Sculptor is an example of a galaxy whose surface brightness profile
deviates from a S\'ersic law at large radii; CGW45, a nucleated dE, on
the other hand, deviates from a S\'ersic profile at small radii. The
differences between the two methods for obtaining $n$ are illustrated
in Fig. \ref{ser1}. A fit to the stellar density profile of Sculptor
yields $n \approx 0.7$; a fit to the growth curve results in $n
\approx 1.0$. The former method yields a very good approximation to
both the growth curve and the density profile within the inner
30~arcmin but fails at larger radii. The latter method gives a
superior fit to the growth curve outside 30~arcmin but fails to
provide an acceptable fit to the density profile. In the case of
CGW45, both methods give $n \approx 1.4$. The central nucleus prevents
both methods from yielding a good fit to the growth curve within the
inner 0.2$''$. Omitting the central 0.2$''$ results in a very good
S\'ersic fit to the density profile.

Choosing between these two methods is clearly a matter of taste. Here,
as in \cite{gg03}, we wish $n$ to reflect the shape of the surface
brightness profile of the bulk of the galaxy and, therefore, opt to
fit a S\'ersic profile to the surface brightness profile evaluated as
a function of equivalent radius.

It is obvious from panel {\bf (b)} of Fig. \ref{LVVI} that for
galaxies fainter than M$_V \sim -14$~mag the relation between
luminosity and $n$ appears to break down. The dwarf spheroidals for
which the S\'ersic exponent $n$ is available, i.e. the Local Group
dSphs, for some of which we have measured $n$ using published star
counts \citep{ih95}, and our new Perseus cluster dSphs, all lie in the
range $n\approx 0.5$ to 1.0, essentially independent of galaxy
luminosity. This $n$ is significantly larger than that predicted by
the scaling relations mentioned before (green curve). It is,
therefore, clear that in the dSph regime, the power-law behaviour of
the M$_V-n$ relation breaks down.

\begin{figure}
\vspace*{8cm} \special{hscale=45 vscale=45 hsize=550 vsize=240
hoffset=-5 voffset=255 angle=-90 psfile="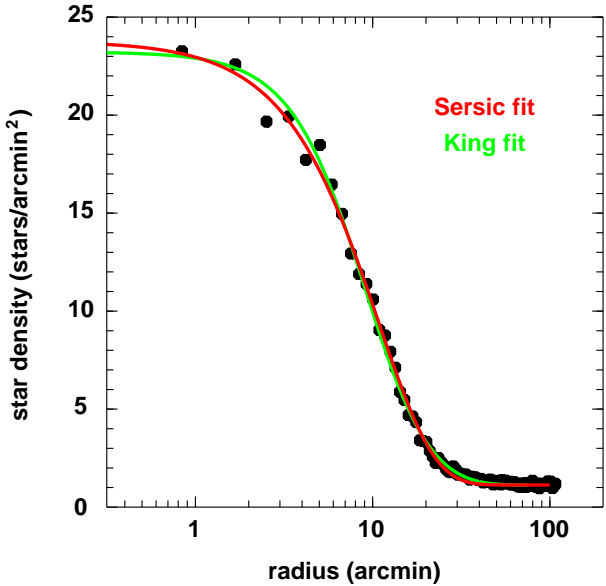"} 
\caption{The radial profile of the projected stellar density
overplotted (black dots) with the best fitting S\'ersic profile (red
line) and King profile (green line). The S\'ersic profile has a shape
parameter $n=0.7$; the concentration of the King profile is $c =
r_{\rm tidal}/r_{\rm core}=4.25$. A constant background of
1.13~arcmin$^{-2}$ was added to the two profiles. \label{ser}}
\end{figure}

From Fig. \ref{ser}, it is clear that a S\'ersic profile with $0.5
\lesssim n \lesssim 1$ is almost indistinguishable from a King model
with a concentration $c = r_{\rm tidal}/r_{\rm core}$ in the range 3
to 10, which is typical for dSphs. In this figure, we plot the radial
profile of projected stellar density of the Sculptor dSph (black data
points), taken from \cite{ih95}, overplotted with the best fitting
S\'ersic profile (red line) and King profile (green line). The
S\'ersic profile has a shape parameter $n=0.7$; the concentration of
the King profile is $c = r_{\rm tidal}/r_{\rm core}=4.25$. Both
3-parameter laws give a very good representation of the data, taking
into account a background density of 1.13~arcmin$^{-2}$. Thus, the
breakdown of the power-law dependence of $n$ on luminosity in the dSph
regime might have been foreseen based on the already known fact that $c
\gtrsim 3$ for dSphs.
%To summarize, the power-law behaviour of the M$_V$ vs. $n$ relation
%breaks down around $n \approx 0.8$ and M$_V \approx -14$~mag, with $n$
%essentially constant below this luminosity.

\cite{sh08} fit the surface brightness profiles of a sample of Local
Volume dSphs, dIrrs, and brighter late-type galaxies with exponentials
and investigate the resulting relation between luminosity and
exponential scale-length, denoted by $h$. They note that the
luminosity-$h$ relation becomes almost flat in the dSph regime. This
is a direct consequence of the M$_V - {\rm R}_{\rm e}$ and M$_V - n$
relations presented here. We generate synthetic S\'ersic surface
brightness profiles using the latter relations. We fit these
synthetic surface brightness profiles with exponentials and thus
constructed a M$_V - h$ relation. This relation is in excellent
agreement with the one found observationally by \cite{sh08}. Between
M$_V=-10$~mag and M$_V=-14$~mag, the exponential scale-length
increases only from 0.25~kpc to 0.5~kpc. This M$_V - h$ relation is
much shallower than the M$_V - {\rm R}_{\rm e}$ relation because of
the increase of $n$ with M$_V$, which makes brighter galaxies more
centrally concentrated than fainter ones.

\subsection{Luminosity vs. central surface brightness}

The M$_V - \mu_{0,V}$ diagram is presented in panel {\bf (c)} of
Fig. \ref{LVVI}. The vertical black line marks the $1\sigma$
background noise level of our HST/ACS images, translated into a V-band
surface brightness. We can only derive reliable surface photometry for
galaxies with a central surface brightness above roughly this
background limit, corresponding to $\mu_{{\rm BG},
  V}=24.5$~mag~arcsec$^{-2}$. For galaxies brighter than M$_V \sim
-14$~mag, the central surface brightness, estimated by extrapolating
the best fitting S\'ersic profile to zero radius, varies as a power of
the luminosity. Only the very brightest cored gEs deviate from this
power law. This underlying unity between Es and dEs was not initially
appreciated since it is not reflected in the luminosity vs. mean and
effective surface brightness diagrams \citep{gg03}. Using the S\'ersic
parameters as fundamental morphological parameters, the assumed
structural dichotomy between Es and dEs has disappeared and the idea
that similar physical processes have governed the evolution of {\em
  all} spheroidal galaxies was put forward.

\begin{figure}
\vspace*{8cm} \special{hscale=50 vscale=50 hsize=550 vsize=240
hoffset=-23 voffset=-133 angle=0 psfile="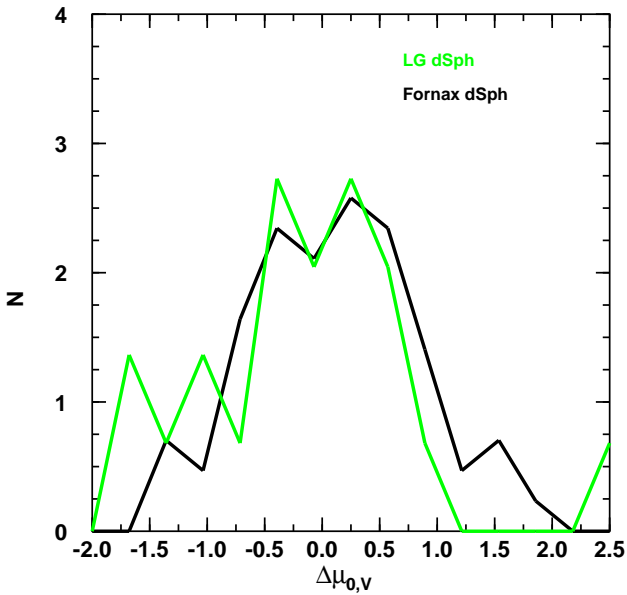"} 
\caption{Distribution of the surface brightness deviation of the Local
  Group dSphs (green curve) and Fornax dSphs (black curve) from the
  mean relation $\mu_{0,V} = (31.86 \pm 0.43) + (0.66 \pm 0.04) \times
  {\rm M}_V$, denoted by $\Delta \mu_{0,V}$. Both distributions have
  the same normalization. Only galaxies fainter than M$_V = -14$~mag
  have been selected for this exercise. The Fornax cluster clearly
  contains a low-surface brightness population that is absent from the
  other datasets. The offset between the means of these distributions
  is about 0.2~mag~arcsec$^{-2}$.  \label{pst}}
\end{figure}

However, for galaxies fainter than M$_V \sim -14$~mag, the slope of
the $\mu_{0,V}-{\rm M}_V$ relation changes significantly. This is a
direct consequence of the near constancy of $n$ in this luminosity
regime, with $n \approx 0.8$ (panel {\bf (b)} of Fig. \ref{LVVI}), and
the observed M$_V - {\rm R}_{\rm e}$ relation (panel {\bf (a)} of
Fig. \ref{LVVI}). At constant $n$, the latter translates into a M$_V -
{\rm r}_0$ relation, with r$_0$ the scale-radius of the S\'ersic
profile, that is completely similar to the M$_V - {\rm R}_{\rm e}$
relation. Given $n=0.8$ and this M$_V - {\rm r}_0$ relation, the
central surface brightness should follow the relation
\begin{equation}
\mu_{0,V} =  29.68 + 0.46\times {\rm M}_V \,\,\,\,\,\, [\rm dSph].
\end{equation}
A linear fit to the dSphs from the Local Group, the Perseus Cluster,
and the Fornax cluster yields
\begin{equation}
\mu_{0,V} = (31.86 \pm 0.43) + (0.66 \pm 0.04) \times {\rm M}_V
\,\,\,\,\,\, [\rm dSph], \label{strl}
\end{equation}
plotted as a orange line in Fig. \ref{LVVI}. This differs signifcantly
from the relation
\begin{equation}
\mu_{0,V} = (43.57 \pm 0.96) + (1.46 \pm 0.06) \times {\rm M}_V\,\,\,\,\,\,\,\,\rm [dE+E]
\end{equation}
fitted to the \cite{gg03} dEs and Es (green line). It is important to
note that the slope change around M$_V=-14$~mag is immediately evident
from the data sets for the dwarf populations of the Fornax cluster and
the Perseus cluster, which cover the transition region in the M$_V -
\mu_{0,V}$ diagram. Hence, within the same environment, dSphs (M$_V
\gtrsim -14$~mag) and dEs/Es (M$_V \lesssim -14$~mag) follow a M$_V -
\mu_{0,V}$ relation with a slope that varies as a function of
luminosity. Early-type galaxies from different environments (the Local
Group, Fornax, Coma, Perseus, \ldots) all fall on the same M$_V -
\mu_{0,V}$ relation. This shows that the slope of this relation is a
strong function of galaxy luminosity but not of environment.

As \cite{mi07} note, their sample includes a population of dSphs that
have much fainter central surface brightnesses than those of the Local
Group. Our HST/ACS imaging, unfortunately, does not go deep enough to
detect such a population in the Perseus cluster. In order to quantify
the effect of this population on the global scaling relation, we
measure the surface brightness deviation of the Fornax dSphs and of
the Local Group and Perseus cluster dSphs from the straight-line
relation given by eq. (\ref{strl}). This deviation of surface
brightness from the mean relation is denoted by $\Delta
\mu_{0,V}$. Only galaxies fainter than M$_V = -14$~mag have been
selected for this exercise. For the Fornax dSphs, the distribution of
this deviation shows a pronounced tail towards large, positive $\Delta
\mu_{0,V}$ (see Fig. \ref{pst}). This is the signature of the
low-surface brightness population. This population is clearly absent
from the Local Group and Perseus dSph datasets. However, the mean
$\Delta \mu_{0,V}$ of the Fornax dSph population differs less than
0.2~mag~arcsec$^{-2}$ from that of the Perseus and Local Group
populations.

Interestingly, dSphs
seem to trace the same M$_V - \mu_{0,V}$ relation as the dIrrs within
the local 10~Mpc volume, which was quantified as
\begin{equation}
\mu_{0,V} = 29.29 + 0.52 \times {\rm M}_V\,\,\,\,\,\,\rm [dIrr]
\end{equation}
by \citet{sh08}. If star formation was somehow switched off in dIrrs
that are initially on this relation, fading of M$_V$ and $\mu_{0,V}$
over time by roughly the same amount (in magnitudes) would result in
exactly such a population of very low surface brightness dSph-like
objects. %The \cite{vdd08} models for isolated dSphs seem to
%interpolate quite naturally between the faint and bright scaling
%relations, with a tendency to stick to the high surface brightness
%edge of the observed dSph M$_V-\mu_{0,V}$ relation.

So, while Es and dEs seem to follow a linear relation between M$_V$
and $\mu_{0,V}$, dSphs deviate significantly from this relation. They
have a higher central surface brightness than predicted by the
extrapolated relation of dEs and Es, albeit with the existence of a
possible cluster population of very low surface brightness dSphs,
which, \cite{mi07} surmise could be a population of tidally heated
dwarf galaxies.

\subsection{Luminosity vs. V$-$I colour} \label{subLVVI}

The V$-$I colours of early-type dwarfs of the Perseus cluster, with
both V- and I-band magnitudes measured within the I-band half-light
radius, are plotted versus V-band absolute magnitude in panel {\bf
  (d)} of Fig. \ref{LVVI}. The same data for Fornax cluster early-type
dwarfs have been taken from \cite{mi07} and plotted in the same
panel. A linear fit to the Fornax cluster M$_V$ vs. V$-$I relation is
overplotted (full red line). The Perseus dwarfs obviously adhere quite
closely to the color-magnitude relation (CMR) defined by the Fornax
dwarfs.

\begin{figure*}
\vspace*{8.5cm}
\special{hscale=50 vscale=50 hsize=720 vsize=450
hoffset=-28 voffset=-130 angle=0 psfile="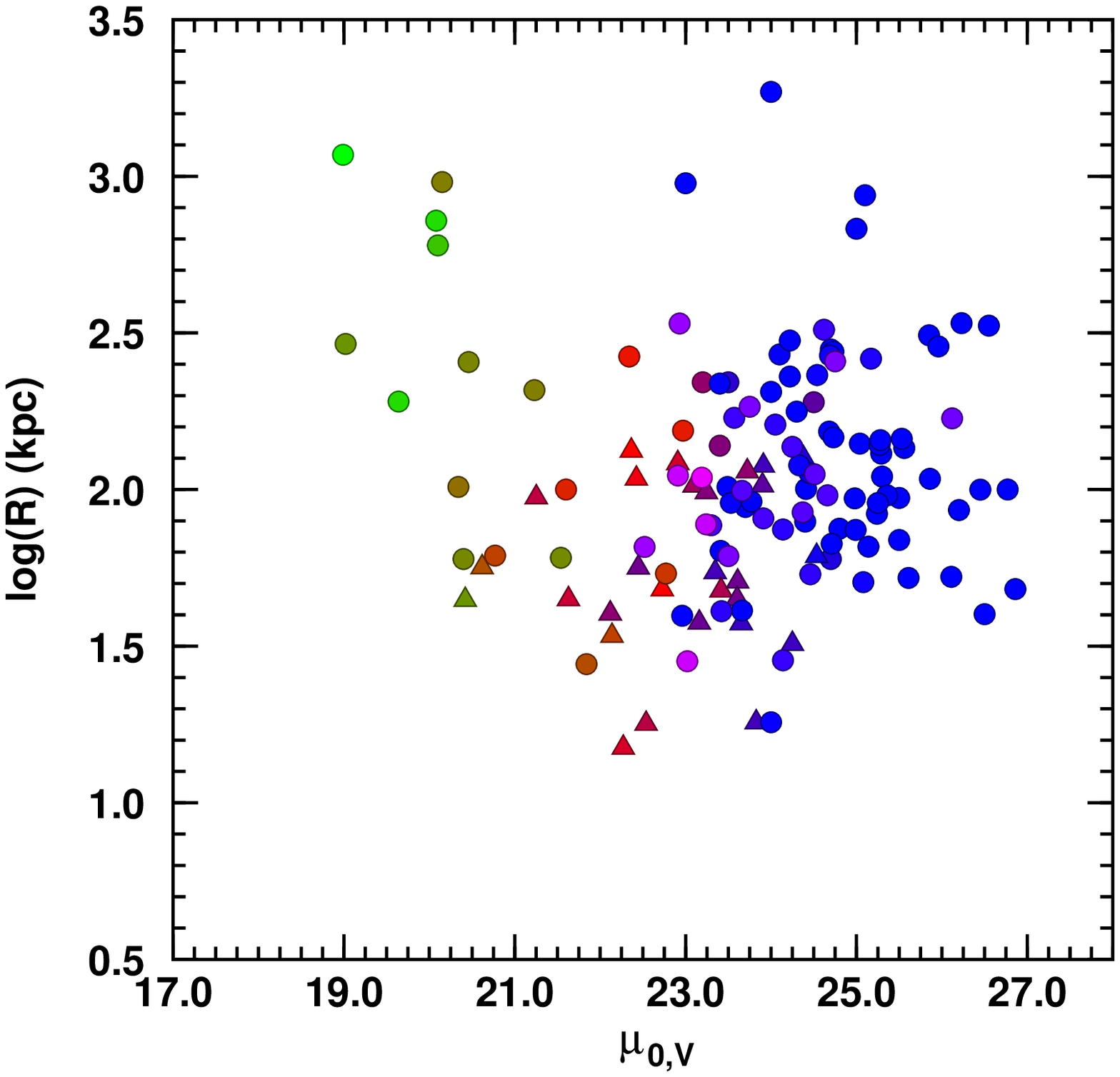"}
\special{hscale=50 vscale=50 hsize=720 vsize=450
hoffset=230 voffset=-130 angle=0 psfile="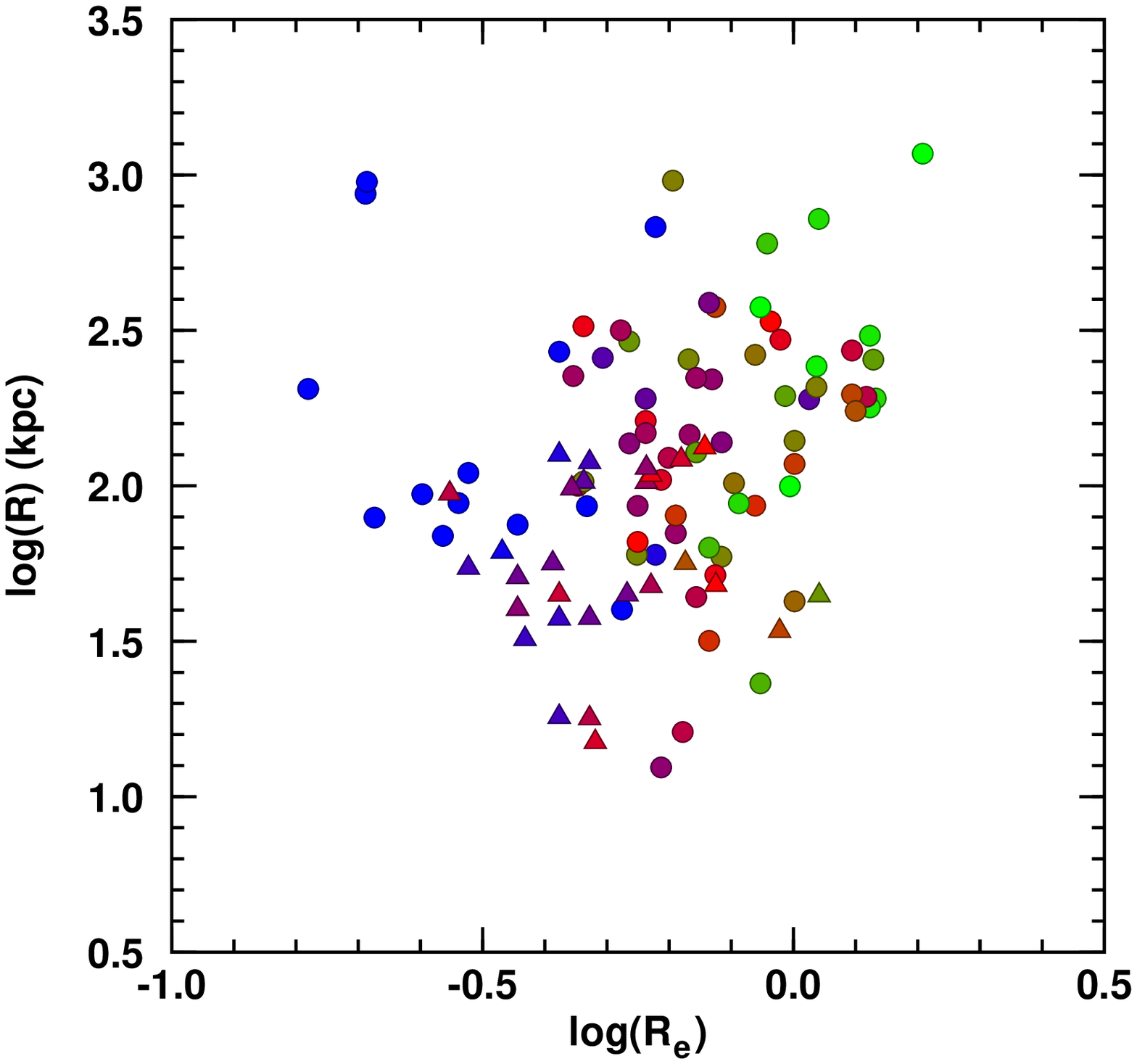"}
\caption{Central surface brightness, $\mu_{0,V}$ (left panel), and
  half-light radius, R$_{\rm e}$ (right panel), as a function of
  projected distance (in kpc), $R$, to the nearest bright galaxy (with
  M$_V<-20$~mag). The triangle data points are the Perseus dSphs. In
  the left panel, the necessary data are available for the dSphs/dEs
  of the Local Group, the Perseus dSphs, and the Fornax cluster
  dSphs/dEs from D05 and Mieske et al. (2007). In the right panel, we
  use data from the dSphs/dEs of the Local Group, the Perseus dSphs,
  the Fornax cluster dEs from D05, and the Antlia dSpsh/dEs from Smith
  Castelli et al. (2008). The data points are color coded according to
  their absolute luminosity, from blue (M$_V=-11$~mag) over red
  (M$_V=-14$~mag) to green (M$_V=-18$~mag). While there is no
  statistically significant correlation between these structural
  parameters and distance to the nearest bright galaxy there is a
  clear correlation with luminosity. \label{pointings1}}
\end{figure*}

In order to investigate the physical nature of the CMR, we convert
the Iron abundance [Fe/H], measured by \cite{m07} for the D05 sample
of cluster and group dEs, into a V$-$I color using an empirical
color-metallicity relation calibrated for old stellar populations,
such as globular clusters \citep{cha90,kw98}. This places the D05
dwarfs essentially along the extension of the \cite{mi07} CMR. The
same exercise can be done for the Local Group dSphs, with the
metallicity taken from the compilation by \cite{ggh03}, with the same
result:~they end up following the \cite{mi07} CMR. This already
suggests that the V$-$I CMR of early-type dwarf galaxies is, in fact,
a luminosity-metallicity relation. As a further test, we convert the
C$-$T$_1$ colors of the Antlia dSphs/dEs, measured by \cite{smc08},
into V$-$I colors using empirical (C$-$T$_1$)$-$[Fe/H] and
[Fe/H]$-$(V$-$I) relations as an intermediate step. This places the
Antlia dSphs/dEs almost exactly on the extension of the CMR of the
Fornax dSphs.

Thus, it appears that the observed V$-$I CMR of early type galaxies,
from dwarfs to giants, is a luminosity-metallicity relation of
galaxies that have stopped forming stars sufficiently long ago for
there being almost no age information left (see also \cite{smc08} and
references therein).% The CMR predicted by the simulations of isolated
%dSphs of \cite{vdd08} tends to be 0.1-0.2~mag bluer than the observed
%CMR. This a consequence of the fact that the model dSphs do not
%efficiently loose their gas and keep forming stars, in several
%distinct star-formation episodes, until the end of the
%simulation. Real dSphs somehow got rid of their gas and have stopped
%forming stars.

\section{Discussion and conclusions} \label{disc}

{\bf We have collected photometric data of dSphs/dEs from different
  galaxy groups and clusters. Analogous to \cite{gyf03}, we quantify
  the galaxy densities of the different environments we are studying
  as the projected density of galaxies brighter than M$_B=-19$~mag
  within the radius $d_5$ that contains five such galaxies, or:
\begin{equation}
\Sigma_5 = \frac{5}{\pi d_5^2}\,{\rm galaxies}\,{\rm Mpc}^{-2}. \label{form}
\end{equation}
The $\log(\Sigma_5)$-values of the different environments from which
we compiled the dataset are listed in Table 1.

\begin{table}
 \centering
 \begin{minipage}{80mm}
  \caption{Central bright-galaxy density, measured by $\log(\Sigma_5)$
    (see eq. (\ref{form})), of the different environments from which
    the dataset was composed.}
  \begin{tabular}{@{}ll@{}}
  \hline
 group/cluster &   $\log(\Sigma_5)$  \\
 \hline
 Local Group    & -0.9 \\
NGC5898 group   & -0.7 \\
 Fornax cluster & 0.5 \\
NGC5044 group   & 0.7 \\
 Antlia cluster & 1.5 \\
Perseus cluster & 1.8 \\
 Coma cluster   & 2.0 \\
 \hline
\end{tabular}
\end{minipage}
\end{table}

The Local Group and the NGC5989 group constitute the sparsest
environments covered by the dataset (the NGC5989 group consists of two
bright ellipticals, NGC5903 and NGC5898, and a few tens of much
fainter galaxies; \citet{gcf92} list only three group members brighter
than M$_B \approx -18$~mag). The Fornax cluster and the NGC5044 group
have comparable, intermediate bright galaxy densities. The Coma,
Perseus, and Antlia clusters have the most extreme central bright
galaxy densities. Obviously, the dataset contains early-type galaxies
from a wide variety of environments.  }

There is considerable uniformity in the photometric properties of
early-type galaxies, from dwarfs to giants. Photometric parameters
quantifying the structure and stellar populations of early-type
galaxies, such as the half-light radius, R$_{\rm e}$ the central
surface brightness $\mu_{0,V}$, the S\'ersic exponent $n$, and V$-$I
color all correlate with galaxy luminosity over a range of more than 6
orders of magnitude in luminosity. The scaling relations involving the
S\'ersic parameters, contrary to previous claims, do not keep a
constant slope over the whole luminosity range. The S\'ersic exponent
$n$ varies with luminosity $L$ as $n \propto L^{0.25 - 0.3}$ for
galaxies brighter than M$_V \approx -14$~mag but scatters around a
constant value within the range $n \approx 0.5 - 1.0$ for fainter
dSphs. This is in agreement with the fact that the surface brightness
profiles of dSphs can be well approximated by King profiles with a
concentration in the range $c \approx 3 - 10$. Central surface
brightness increases with luminosity until the formation of the very
brightest, cored ellipticals. {\bf The cores in the most luminous
  ellipticals are thought to result from the partial evacuation of the
  nuclear region by coalescing black holes (see GG03 and references
  therein).} At M$_V \approx -14$~mag, the slope of the M$_V -
\mu_{0,V}$ changes abruptly. We have shown that the M$_V$ vs. V$-$I is
essentially a metallicity-luminosity relation of old stellar
populations, keeping the same slope over the whole luminosity range
investigated here.

Clearly, the absolute magnitude M$_V \approx -14$~mag is not just an
arbitrary divide between dSphs and dEs. The rather abrupt changes in
the slopes of some of the photometric scaling relations suggest that
below and above this luminosity, different physical processes dominate
the evolution of early type galaxies. The near-independence of these
scaling relations with respect to environment and the physical
differences between dSphs and dEs will be investigated theoretically
using N-body/SPH-models in another paper in this series (De Rijcke,
Valcke, Conselice, Penny, Held, in prep.). In a sense, the divide
between dEs and Es, which has historically been placed at M$_V \approx
-19$~mag seems more arbitrary since the behaviour of the basic
parameters describing the shapes of the surface brightness profiles as
a function of luminosity (i.e. the S\'ersic parameters) does not
change. A comparison of numerical simulations with observed scaling
relations suggests that the luminosity dependence of the S\'ersic
parameters is due to fact that the effects of supernova feedback
become more important as galactic gravitational potential wells become
more shallow for lower galaxy masses \citep{c01,ny04,vdd08}. This
causes dEs to be more diffuse and to have stars orbiting with lower
velocities than predicted by the extrapolated relations for Es
\citep{h92,gg03,mg05,de05}.

{\bf These scaling relations are amazingly insensitive to (local)
  environment. In Fig. \ref{pointings1}, we show the central surface
  brightness, $\mu_{0,V}$ (left panel), and half-light radius, R$_{\rm
    e}$ (right panel), as a function of projected distance (in kpc),
  $R$, to the nearest bright galaxy (with M$_V<-20$~mag). The triangle
  data points are the Perseus dSphs. In the left panel, the necessary
  data are available for the dSphs/dEs of the Local Group, the
  Perseus dSphs, and the Fornax cluster dSphs/dEs from D05 and Mieske et
  al. (2007). In the right panel, we use data from the dSphs/dEs of
  the Local Group, the Perseus dSphs, the Fornax cluster dEs from D05,
  and the Antlia dSpsh/dEs from Smith Castelli et al. (2008). The data
  points are color coded according to their absolute luminosity, from
  blue (M$_V=-11$~mag) over red (M$_V=-14$~mag) to green
  (M$_V=-18$~mag). While there is no statistically significant
  correlation between these structural parameters and distance to the
  nearest bright galaxy there is a strong correlation with luminosity.
  
  We have also plotted these quantities as a function of distance to
  the cluster or group center divided by the virial radius
  R$_{200}$. Again, no correlation with position becomes apparent. The
  galaxies with central surface brightness
  $\mu_{0,V} \approx 23$~mag~arcsec$^{-2}$, where the slope of the
  M$_V-\mu_{0,V}$ relation changes, have nearest neighbor distances
  scattering between 20 and 300~kpc. However, they all have
  luminosities M$_V \approx -14$~mag. From this exercise, we can
  conclude that the photometric scaling relations presented in
  Fig. \ref{LVVI} are not a consequence of environmental segregation,
  with fainter galaxies preferentially located close to a bright
  galaxy.

  \citet{pac08} study the structural parameters such as the
  concentration index, Petrosian radius, velocity dispersion, u$-$r
  colour, \ldots of a volume-limited sample of 49,571 galaxies
  extracted from the SDSS. They show that these parameters are almost
  independent of large-scale density and neighbor separation unless
  the latter is smaller than about one-tenth of the bright neighbor's
  virial radius, i.e., of the order of a few tens of kpc (their
  Figs. 7 and 9). All galaxies presented in the present paper have
  projected distances between 0.1 and 1 virial radii away from their
  nearest bright neighbor and, as \cite{pac08}, we observe no
  significant relation between nearest-neighbor distance or
  environment density and structural properties in this regime. This
  also corroborates the results of \cite{w08} who study a sample of
  galaxies selected from the SDSS DR4 and find that the structural
  properties of early-type satellite galaxies are very similar to
  early-type central galaxies. Hence, the structure of early-type
  galaxies is not significantly affected by environmental effects.

} 

The most obvious environmental effect appears to be the population of
low surface brightness dSphs discovered by \cite{mi07} in the Fornax
cluster. Where an environmental influence is clearly discernible, it
has only mild effects on the scaling relations. The presence of a
low-surface brightness population of Fornax dSphs doesn't affect the
global scaling relation appreciably. At a given luminosity, the Fornax
dSph population is on average 0.2~mag~arcsec$^{-2}$ fainter than the
Perseus and Local Group dSph populations. The M31 companions with
tidal extensions or distorsions \citep{s07,l07} are not displaced from
the general scaling relations. This may indicate that dSphs
intrinsically have high enough M/L to survive unscathed (see
\cite{pe08}) or that dSphs with too low M/L have been destroyed and
only those with high M/L survive to the present day.

\section*{Acknowledgments}

SDR wishes to thank Philippe Prugniel and Mina Koleva for their
hospitality and for the stimulating discussions while visiting CRAL
Lyon Observatory during the course of this work. CJC and SJP
acknowledge support from STFC. SDR is a Postdoctoral Fellow of the
Fund for Scientific Research - Flanders (Belgium)(F.W.O). SV is a PhD
Fellow of the Fund for Scientific Research - Flanders
(Belgium)(F.W.O). This research has made use of the NASA/IPAC
Extragalactic Database (NED) which is operated by the Jet Propulsion
Laboratory, California Institute of Technology, under contract with
the National Aeronautics and Space Administration.

\bsp

\label{lastpage}


\begin{thebibliography}{99}
\bibitem[Bender et al. (1989)]{b89} Bender R., Surma P., D\"obereiner
S., M\"ollenhoff C., Madejsky R., A\&A, 217, 35
%\bibitem[Binggeli \& Popescu (1995)]{bp95} Binggeli B. \& Popescu
%C. C., 1995, A\&A, 298, 63
\bibitem[Blitz \& Robishaw (2000)]{br00} Blitz L. \& Robishaw T.,
2000, ApJ, 541, 675
\bibitem[Buyle et al. (2005)]{b05} Buyle P., De Rijcke S.,
Michielsen D., Baes M., Dejonghe H., 2005, MNRAS, 360, 853
%\bibitem[Bouchard, Jerjen, Da Costa (2007)]{bo07} Bouchard A., Jerjen
%  H., Da Costa, G. S., 2007, AJ, 133, 261
\bibitem[Caldwell (1999)]{c99} Caldwell N., 1999, AJ, 118, 1230
\bibitem[Carraro et al. (2001)]{c01} Carraro G., Chiosi C., Girardi
L., Lia C., 2001, MNRAS, 335, 335
\bibitem[Conselice, Gallagher, Wyse (2003)]{cgw03} Conselice C. J.,
Gallagher, J. S., {\sc iii}, Wyse R. F. G., 2003, AJ, 125, 66
\bibitem[Conselice et al. (2003)]{cogw03} Conselice, C.J., O'Neil, K.,
  Gallagher, J.S., Wyse, R.F.G. 2003, ApJ, 591, 167
\bibitem[Conselice (2003b)]{c03} Conselice C. J., 2003, ApJS, 147, 1
\bibitem[Couture, Harris, Allwright (1990)]{cha90} Couture J., Harris
W. E., Allwright J. W. B., 1990, ApJS, 73, 671
%\bibitem[De Rijcke et al. (2001)]{de01} De Rijcke S., Dejonghe H.,
%Zeilinger W. W., Hau G. K. T., 2001, ApJ, 559, L21
\bibitem[De Rijcke et al. (2003)]{de03} De Rijcke S., Zeilinger W. W.,
Dejonghe H., Hau G. K. T., 2003, MNRAS, 339, 225
\bibitem[De Rijcke et al. (2005)]{de05} De Rijcke S., Michielsen D.,
Dejonghe H., Zeilinger W. W., Hau G. K. T., 2005, MNRAS, 360, 853 (D05)
\bibitem[De Rijcke et al. (2006)]{de06} De Rijcke S., Prugniel P.,
Simien F., Dejonghe H., 2006, MNRAS, 369, 1321
\bibitem[De Rijcke et al. (2007)]{de07} De Rijcke S., Zeilinger W. W.,
Hau G. K. T., Prugniel P., Dejonghe H., 2007, ApJ, 659, 1172
%\bibitem[Dolphin (2002)]{dae02} Dolphin A. E., 2001, MNRAS,
%332, 91
%\bibitem[Dolphin et al. (2005)]{dwsh05} Dolphin A. E., Weisz D. R.,
%Skillman E. D., Holtzman J. A., 2005, 2005astro.ph..6430, Invited
%Review at the meeting "Resolved Stellar Populations", held in Cancun,
%Mexico, 18-22 April 2005
%\bibitem[Gallart et al. (2001)]{g01} Gallart C., Mart\'{\i}nez-Delgado
%D., G\'omez-Flechoso M. A., Mateo M., AJ, 121, 2572
\bibitem[Goto et al. (2003)]{gyf03} Goto T., Yamauchi C., Fujita Y.,
  Okamura S., Sekiguchi M., Smail I., Bernardi M., Gomez P. L., 2003,
  MNRAS, 346, 601
\bibitem[Gourgoulhon, Chamaraux, Fouqu\'e (1992)]{gcf92} Gourgoulhon
  E., Chamaraux P., Fouqu\'e P., A\&A, 255, 69
\bibitem[Graham \& Guzm\'an (2003)]{gg03} Graham A. W. \& Guzm\'an R.,
2003, AJ, 126, 1787 (GG03)
\bibitem[Graham \& Worley (2008)]{gw08} Graham A. W. \& Worley C. C.,
  2008, accepted for publication in MNRAS, 2008arXiv0805.3565
\bibitem[Grebel, Gallagher, Harbeck (2003)]{ggh03} Grebel E. K.,
Gallagher J. S. {\sc iii}, Harbeck D., 2003, AJ, 125, 1926
\bibitem[Held et al. (1992)]{h92} Held E. V., de Zeeuw T., Mould J.,
  Picard A., 1992, AJ, 103, 851
%\bibitem[Hurley-Keller, Mateo, Nemec (1998)]{hmn98} Hurley-Keller D.,
%Mateo M., Nemec J., 1998, AJ, 115, 1840
\bibitem[Irwin \& Hatzidimitriou (1995)]{ih95} Irwin M. \&
Hatzidimitriou D., 1995,
\bibitem[Jerjen \& Binggeli (1997)]{jb97} Jerjen H. \& Binggeli B.,
1997, The Nature of Elliptical Galaxies, 2nd Stromlo Symposium. ASP
Conference Series, Vol. 116, 1997, eds. M. Arnaboldi, G. S. Da Costa,
and P. Saha (1997), p.239
\bibitem[Jerjen (2003)]{j03} Jerjen H., 2003, A\&A, 398, 63
\bibitem[Johnston, Spergel, Hernquist (1995)]{jsh95} Johnston K. V.,
Spergel D. N., Hernquist L., 1995, ApJ, 451, 598
\bibitem[Kleyna et al. (2005)]{k05} Kleyna J. T., Wilkinson
M. I. Evans N. W., Gilmore G., 2005, ApJ, 630, L141
\bibitem[Kundu \& Whitmore (1998)]{kw98} Kundu A. \& Whitmore B. C.,
1998, AJ, 116, 2841
\bibitem[Lewis et al. (2007)]{l07} Lewis G. F., Ibata R. A., Chapman
S. C., McConnachie A., Irwin M. J., Tolstoy, E., Tanvir N. R., 2007,
MNRAS, 375, 1364
\bibitem[Lisker et al. (2006)]{l06} Lisker T., Glatt K., Westera P.,
Grebel E. K., 2006, AJ, 132, 2432
\bibitem[{\L}okas (2002)]{l02} {\L}okas E. L., 2002, MNRAS, 333, 697
\bibitem[Marcolini, Brighenti, D'Ercole (2003)]{mbe03} Marcolini A.,
  Brighenti F., D'Ercole A., 2003, MNRAS, 345, 1329
\bibitem[Marcolini et al. (2006)]{m06} Marcolini A., D'Ercole A.,
Brighenti F., Recchi S., 2006, MNRAS, 371, 64
\bibitem[Martin, de Jong, Rix (2008)]{mdr08} Martin N. F., de Jong
  T. A., Rix H.-W., 2008, accepted for publication in ApJ,
  2008arXiv:0805.2945
\bibitem[Mateo et al. (1996)]{m96} Mateo M., Mirabal N., Udalski A.,
Szymanski M., Kaluzny J., Kubiak M., Krzeminski W., Stanek K. Z.,
1996, ApJ, 458, L13
\bibitem[Mateo (1998)]{m98} Mateo M. L., 1998, ARA\&A, 36, 435
\bibitem[Mateo, Olszewski, Walker (2008)]{m08} Mateo M., Olszewski
E. W., Walker M. G., 2008, ApJ, 675, 201
\bibitem[Matkovi\'c \& Guzm\'an (2005)]{mg05} Matkovi\'c A. \&
Guzm\'an R., 2005, MNRAS, 362, 289
%\bibitem[Mayer et al. (2006)]{ma06} Mayer L., Mastropietro C., Wadsley
%J., Stadel J., Moore B., 2006, MNRAS, 369, 1021
\bibitem[McConnachie \& Irwin (2006)]{mi06} McConnachie A. W. \& Irwin
M. J., 2006, MNRAS, 365, 1263
\bibitem[McConnachie, Arimoto, Irwin (2007)]{mai07} McConnachie A. W.,
Arimoto N., Irwin M., 2007, MNRAS, 379, 379
%\bibitem[McGaugh et al. (2000)]{m02} McGaugh S. S., Schombert J. M.,
%Bothun G. D., de Block W. J. G., 2000, ApJ, 533, L99
\bibitem[Michielsen et al. (2007)]{m07} Michielsen D. Koleva M.,
Prugniel P., Zeilinger W. W., De Rijcke S., Dejonghe H., Pasquali A.,
Ferreras I., Debattista V. P., 2007, ApJ, 670, L101
\bibitem[Mieske et al. (2007)]{mi07} Mieske S., Hilker M., Infante L,
Mendes de Oliviera C., 2007, A\&A, 463, 503
\bibitem[Nagashima \& Yoshii (2004)]{ny04} Nagashima M. \& Yoshii Y.,
2004, ApJ, 610, 23
%\bibitem[Patterson \& Thuan (1996)]{pt96} Patterson R. J. \& Thuan
%T. X., 1996, ApJS, 107, 103
\bibitem[Park \& Choi (2008)]{pac08} Park C. \& Choi Y.-Y., 2008,
  accepted by ApJ, arXiv:0809.2156
\bibitem[Peletier (1993)]{p93} Peletier R. F., 1993, A\&A, 271, 51
\bibitem[Penny \& Conselice (2008)]{pc08} Penny S. J. \& Conselice
C. J., 2008, MNRAS, 383, 247
\bibitem[Penny et al. (2008)]{pe08} Penny S. J., Conselice C. J., De
  Rijcke S., Held E. V., 2008, submitted to MNRAS
\bibitem[Prugniel \& Simien (1997)]{ps97} Prugniel P. \& Simien F.,
A\&A, 321, 111
\bibitem[Ricotti \& Gnedin (2005)]{ri05} Ricotti M. \& Gnedin N. Y.,
  2005, ApJ, 629, 259
\bibitem[Saviane, Held, Piotto (1996)]{sv96} Saviane I., Held E. V.,
Piotto G., 1996, A\&A, 315, 40
\bibitem[Schlegel, Finkbeiner, Davis (1998)]{sch98} Schlegel D. J.,
Finkbeiner D. P., Davis M., 1998, ApJ, 500, 525
\bibitem[S\'egall et al. (2007)]{s07} S\'egall M., Ibata R. A., Irwin
M. J., Martin N. F., Chapman S., 2007, MNRAS, 375, 831
\bibitem[Sharina et al. (2008)]{sh08} Sharina M. E., Karachentsev
I. D., Dolphin A. E., Karachentseva V. E., Tully R. B., Karataeva
G. M., Makarov D. I., Makarova L. N., Sakai S., Shaya E. J., Nikolaev
E. Y., Kuznetsov A. N., 2008, MNRAS, 384, 1544 
%\bibitem[Simon \& Geha (2007)]{sg07} Simon J. D. \& Geha M., 2007,
%ApJ, 670, 313
\bibitem[Sirianni et al. (2005)]{si05} Sirianni M., Jee M. J.,
Ben\'{\i}tez N., Blakeslee J. P., Martel A. R., Meurer G., Clampin M.,
De Marchi G., Ford H. C., Gilliland R., Hartig G. F., Illingworth
G. D., Mack J., McCann W. J., 2005, PASP, 117, 1049
\bibitem[Smith Castelli et al. (2008)]{smc08} Smith Castelli A. V.,
  Bassino L. P., Richtler T., Cellone S. A., Aruta C., Infante L.,
  2008, MNRAS, 386, 2311
\bibitem[Struble \& Rood (1999)]{Stublerood99} Struble M.F., Rood
H.J., 1999, ApJS, 125, 36
\bibitem[Tonry, Dressler, Blakeslee (2001)]{t01} Tonry J. L., Dressler
  A., Blakeslee J. P., et al., 2001, ApJ, 546, 681
%\bibitem[Sung et al. (1998)]{su98} Sung E.-C., Han C., Ryden B. S.,
%Patterson R. J., Chun M.-S., Kim H.0-L., Lee W.-B., Kim D.-J., 1998,
%ApJ, 505, 199
%\bibitem[Thomas et al. (2005)]{th05} Thomas D., Maraston C., Bender
%R., Mendes de Oliviera C., 2005, ApJ, 621, 673
%\bibitem[Tully \& Trentham (2008)]{tt08} Tully R. B. \& Trentham N.,
%2008, AJ, 135, 1488
\bibitem[Valcke, De Rijcke, Dejonghe (2008)]{vdd08} Valcke S., De
  Rijcke S., Dejonghe H., 2008, accepted for publication in MNRAS
%\bibitem[van Eymeren et al. (2007)]{v07} Van Eymeren J., Bomans D. J.,
%Weis K., Dettmar R.-J., 2007, A\&A, 474, 67
%\bibitem[van Zee, Skillman, Haynes (2004)]{vz04} van Zee L., Skillman
%E. D., Haynes, M. P., 2004, AJ, 128, 121
\bibitem[van Zee, Barton, Skillman (2004)]{vzb04} van Zee L., Barton
E. J., Skillman E. D., 2004, AJ, 128, 2797
\bibitem[Vazdekis et al. (1996)]{vz96} Vazdekis A., Casuso E.,
  Peletier R. F., Beckman J. E., 1996, ApJS, 106, 307
\bibitem[Weinmann et al. (2008)]{w08} Weinmann S., M., Kauffmann G.,
  van den Bosch F. C., Pasquali A., McIntosh D., H., Mo H., Yang X.,
  Guo Y., 2008, submitted to MNRAS, 2008arXiv0809.2283
\bibitem[Young et al. (2007)]{y07} Young L. M., Skillman E. D., Weisz
D. R., Dolphin A. E., 2007, ApJ, 659, 331
\bibitem[Zucker et al. (2007)]{z07} Zucker D. B., Kniazev A.  Y.,
Mart\'{\i}nez-Delgado D., Bell E. F., Rix H.-W., et al., 2007, ApJ,
659, L21
\end{thebibliography}
\end{document}